\documentclass[12pt]{iopart}

\usepackage{graphicx}

\begin{document}

\title{Size effects in statistical fracture}

\author{Mikko J. Alava$^{1}$}
\address{$^{1}$ Laboratory of Physics, Helsinki University of
Technology, FIN-02015 HUT}
\author{Phani K. V. V. Nukala$^2$}
     \address{$^2$ Computer Science and Mathematic Division,
Oak Ridge National Laboratory, Oak Ridge, TN 37831-6359, USA}
\author{Stefano Zapperi$^{3,4}$}
    \address{$^3$ INFM-CNR, S3, Dipartimento di Fisica, Universit\`a di Modena e Reggio
Emilia, via Campi 213/A, I-41100, Modena, Italy}
\address{$^4$ ISI Foundation, Viale S. Severo 65, 10133 Torino,
Italy }

\begin{abstract}
We review statistical theories and numerical methods employed to
consider the sample size dependence of the failure strength
distribution of disordered materials. We first overview the
analytical predictions of extreme value statistics and fiber bundle
models and discuss their limitations. Next, we review energetic and
geometric approaches to fracture size effects for specimens with a
flaw. Finally, we overview the numerical simulations of lattice
models and compare with theoretical models.
\end{abstract}


\maketitle

\date{\today}
\section{Introduction}

Understanding how materials break is a fundamental problem of
science and engineering. The difficulties stem from the non-trivial
dependence of the fracture strength on the characteristic
lengthscales of the samples. This was already noted by Leonardo da
Vinci, who measured the carrying-capacity of metal wires of varying
length \cite{leonardo40}. He observed that the longer the wire, the
less weight it could sustain. The reason for this behavior is rooted
in the disorder present in the material. If a sample can be divided
into non-interacting subvolumes, then the strength is dominated by
the weakest one and its distribution could in principle be computed
using extreme value statistics \cite{gumbel,weibull39}. Longer wires
are more likely to possess weak parts and are thus bound to fail at
smaller loads on average. The quantitative understanding of this
statistical size effect is difficult and the low tails of the
strength probability distribution are not easy to sample
experimentally. In addition, microcracks interact by long-range
stress fields. Hence, the assumption made to use extreme value
statistics, that of independent subvolumes, is difficult to justify
in practice. These problems are particularly compelling in
quasi-brittle materials, such as concrete and many other composites,
where sample failure is preceded by significant damage accumulation
\cite{vanmierbook}.

The typical materials science question is to study the size effect
with a pre-existing flaw, {\it a notch}. Failure in quasi-brittle
materials is then determined by the competition between
deterministic effects, due to the stress enhancement created by
notch, and the damage accumulation around the defect due to the
stress concentration \cite{bazantbook,bazant99}. The effect of
disorder is then treated in an effective medium sense by defining a
Fracture Process Zone (FPZ) around the crack tip. Starting from
these observations, several theoretical formulations based on linear
elastic fracture mechanics (LEFM) have been proposed in the
literature and compared with experiments
\cite{bazantbook,bazant99,hu92,bazant04b,morel00,morel02,alava08a}.
The problem is how to extend LEFM when disorder is strong and can
not be treated as a small perturbation on the homogeneous solution.

The statistical physics approach to fracture is based on simple
lattice models, which allow for a relatively simple description of
disorder and elasticity. The models are sometimes amenable to
analytical solutions, and usually are simulated numerically. In the
simplest approximation, elastic interactions are replaced by a load
transfer rule which is applied when elements fail or get damage.
These "fiber bundle models" can be solved exactly in some cases and
can thus provide a useful guidance for the simulations of more
realistic models in which the elastic medium is represented by a
network of springs or beams. In this case, the local displacements
can then be found by standard method for solving coupled linear
equations. Disorder is modeled for instance by imposing random
failure thresholds on the springs or by removing a fraction of the
links. The lattice is loaded imposing appropriate boundary
conditions and the fracture process can be followed step by step, in
a series of quasi-equilibria. The cornerstone in this respect has
been for the last twenty years the Random Fuse Model (RFM)
\cite{deArcangelis85}, a lattice model of the fracture of solid
materials in which as a further key simplification vectorial
elasticity has been substituted with a scalar field.

In this paper, we review the statistical physics approaches used to
understand size effects in fracture. We first discuss the
traditional weakest-link approaches and the solution of fiber bundle
models. Next, we consider size effects arising from energetic
considerations when a single dominating crack is present, and when
its growth may be of importance. Finally, we discuss the results of
lattice models for fracture and compare the results obtained with
the theoretical arguments discussed previously.

\section{Size effect from extreme value statistics}

\subsection{The weakest link model \label{sec:weak}}

The key theoretical concepts needed to understand the distribution
of fracture strength and the associated size effects in
heterogeneous media date back to the pioneering work of Gumbel
\cite{gumbel}, Pierce \cite{pierce26} and Weibull \cite{weibull39}
on the statistics of extremes. The general idea stems from a weakest
link argument: the failure strength of an extended object is ruled
by its weakest local subvolume. For a disordered system, the larger
the sample the easiest it is to find a weak region.

The easiest way to illustrate this concept is based on a
one-dimensional model. Consider a chain composed by $N$ elastic
links that can sustain at most a stress $s_i$ without breaking. We
assume that the threshold stresses $s_i$ are independent random
variables distributed according to a probability density function
$p(s_i)$ and a corresponding cumulative distribution ${\cal
P}(s_i<\sigma) = P(\sigma)=\int_0^\sigma p(x) dx$. The chain will
fail as soon as one of the links, the weakest one, fails. Hence, the
probability that a chain fails at a stress larger than $\sigma$ is
equal to the probability that the {\it minimum} of $s_i$ is larger
than $\sigma$, which is just the probability that all the links have
thresholds larger than $\sigma$. This condition can be expressed in
mathematical terms for  a chain of $N$ links as
\begin{equation}
1-P_N(\sigma) = \prod_{i=1}^N {\cal P}(s_i>\sigma) =(1-P(\sigma))^N,
\label{eq:extreme}
\end{equation}
where $P_N(\sigma)$ is the probability that the failure stress is
smaller than $\sigma$. In the large $N$ limit,
Eq.~(\ref{eq:extreme}) can be approximated as
\begin{equation}
P_N(\sigma)\simeq 1-\exp [-NP(\sigma)].
\end{equation}
Under some broad assumptions on the tail of $P(\sigma)$, it can be
shown that $P_N(\sigma)$ converges to a stable asymptotic form (i.e.
a distribution with a shape that does not depend on $N$). In more
technical terms a stable distribution is invariant under
Eq.~(\ref{eq:extreme}) provided the variable is linearly transformed
according to $\sigma_N=a_N\sigma+b_N$. For large $N$, the statistics
is dominated by the low value tail of the distribution $P(\sigma)$.
If we assume it to scale close to $\sigma=0$ as $P(\sigma) \simeq
(\sigma/\sigma_0)^\mu$, where $\sigma_0$ is a characteristic stress.
Under this assumption, we obtain
\begin{equation}
P_N(\sigma) \simeq 1-\exp [-N(\sigma/\sigma_0)^\mu],
\label{eq:weibull}
\end{equation}
which is invariant under Eq.~(\ref{eq:extreme}) if we transform the
variable $\sigma$ with $a_N=N^{-1/\mu}$ and $b_N=0$.

Eq.~(\ref{eq:weibull}) is the celebrated Weibull distribution
\cite{weibull39}, predicting that the average strength of a chain
composed by $N$ links decreases as
\begin{equation}
\langle \sigma_N \rangle \propto N^{-1/\mu}.
\end{equation}
The Weibull distribution represents still today the main tool used
to analyze failure statistics in various materials, although the
validity of its underlying assumptions is in general difficult to
demonstrate. Real samples can not generally be schematized as a
chain of independent elements with random failure thresholds. In
many cases, such as in quasibrittle materials, the sample does not
even fracture at once but sustains a considerable amount of damage
before failure. Furthermore, long-range elastic interactions could
correlate different regions of the sample invalidating the
assumptions used to derive the Weibull statistics. That the
interactions and microcrack grow would be irrelevant at sufficiently
large scale, leading to the recovery of the Weibull distribution,
has never been proven rigorously.

\subsection{The largest crack model}

It would be desirable to relate the failure statistics to some
geometrical characteristics of the microstructure of a material,
going beyond a simple description of a sample as a collection of
regions with different random strengths. An attempt in this
direction was proposed by Freudenthal \cite{freudenthal68} and then
rederived later by various authors for different models
\cite{chakrabarti,duxbury86,duxbury87}. The idea is to schematize
the microstructure as a set of independent linear cracks of length
$a_i$, distributed according to given probability density function
$p(a_i)$. If the cracks are sufficiently diluted, we can treat them
as completely isolated and assess their stability by the simple
energetic argument due to Griffith. The idea is establish the
conditions under which crack growth becomes energetically favorable.
Considering for simplicity a two dimensional geometry, the elastic
energy released by a crack of length $a$ is given by
\begin{equation}
{\cal E}_{el} = -\frac{\pi \sigma^2 a^2}{2E},
\end{equation}
where $\sigma$ is the applied stress and $E$ is the Young modulus.
Forming a crack involves the creation of crack surface with an
energy cost which in the ideal case would be related the rupture of
atomic bonds,
\begin{equation}
{\cal E}_{surf}=2 a G,
\end{equation}
where $G$ is the fracture toughness. The crack becomes
unstable when the total energy decreases
\begin{equation}
\frac{d {\cal E}}{ d a} =  -\pi \sigma^2 a/E +2 G <0,
\end{equation}
which implies that a crack of length $a$ becomes unstable when the
stress is equal to
\begin{equation}
\sigma_c=\sqrt{\frac{2EG}{\pi a}}.
\label{eq:griffith}
\end{equation}

If we have a collection of $N$ independent random cracks in a volume $V$, we can assume that
the sample will fail when least stable one will become unstable. The problem
thus reduces to find the distribution of the largest crack. As
discussed in Sec.~\ref{sec:weak},
the cumulative distribution is given by
\begin{equation}
P_N(a)\simeq \exp -[\rho V P(a)],
\label{eq:pna}
\end{equation}
where $\rho\equiv N/V$ is the crack density and $P(a)$ is the
cumulative distribution associated with the crack density
distribution $p(a)$. The asymptotic behavior of Eq.~(\ref{eq:pna})
is ruled by the large-value tail of $P(a)$. It can be shown that if
the tail decays faster than an exponential, Eq.~(\ref{eq:pna})
converges to the Gumbel distribution \cite{gumbel} for large $N$
\begin{equation}
P_N(a) =\exp [-\rho V\exp(-a/a_c)],
\label{eq:gumbel}
\end{equation}
where $a_c$ is the characteristic scale of the crack length
distribution. Combining Eqs.~(\ref{eq:gumbel}) and
(\ref{eq:griffith}), one can derive the strength distribution as
\begin{equation}
P_N(\sigma) =1-\exp \left[-\rho V \exp\left(-\frac{\sigma_0}{\sigma}\right)^{2}\right],
\label{eq:gumbel2}
\end{equation}
where $\sigma_0\equiv 2EG/\pi a_c$. Eq.~(\ref{eq:gumbel2}) predicts
a size effect for the average strength of a type logarithmic in
size:
\begin{equation}
\langle \sigma_V \rangle \simeq A/(B+C\log(V)).
\end{equation}

\subsection{Interacting cracks: fiber bundle models}

The main shortcoming of extreme value statistics is the fact that
local failure stresses are considered independent. This is difficult
to justify in practice since cracks induce long-range interactions
that may correlate stresses in different regions of the sample. A
simple way to analyze the role of interaction in the fracture of
disordered media is represented by the study of fiber bundle models
\cite{pierce26,daniels45}. These models consider a set of brittle
fibers, with random failure thresholds, loaded in parallel. When a
fiber breaks its load is redistributed to the other fibers according
to a prescribed rule. The simplest possibility is the case of an
equal load sharing (ELS), in which each intact fiber carries the
same fraction of the load. This case represents a sort of mean-field
approximation and allows for a complete analytic treatment
\cite{daniels45,hemmer92,hansen94,sornette92,kloster97}. At the
other extreme lies the local load sharing model (LLS) where the load
of a failed fiber is redistributed to the intact neighboring fibers
\cite{harlow78,smith80,smith81,beyerlein96,phoenix97,phoenix92,curtin91,curtin93,zhou95,leath94,zhang96}

We consider first the case of ELS fiber bundles, in which $N$ fibers
of unitary Young modulus $E=1$ are subject to an uniaxial load $F$.
Each fiber $i$ obeys linear elasticity up to a critical load $x_i$,
which is randomly distributed according to a distribution $p(x)$.
When the load on a fiber exceeds $x_i$, the fiber is removed. Due to
the ELS rule, when $n$ fibers are present each of them carries a
load $F_i = F/n$ and consequently a strain $\epsilon=F/n$. The
constitutive law for ELS fiber bundles can be easily be obtained
from a self-consistent argument. At a given load $F$, the number of
intact fibers is given by
\begin{equation}
n= N\left(1-\int^{F/n}_0 p(x) dx \right).
\label{eq:fbm_sc}
\end{equation}
Rewriting Eq.~(\ref{eq:fbm_sc}) as a function of the strain, we obtain the
constitutive law
\begin{equation}
 \label{eq:constone}
  \frac{F}{N} = \epsilon (1-P(\epsilon)),
\end{equation}
where $P(x)$ is the cumulative distribution obtained from $p(x)$.
For simplicity we rewrite Eq.~(\ref{eq:constone}) as $f=\epsilon (1-P(\epsilon))$
where $f/\equiv F/N$ is the load per fiber. Failure corresponds to the maximum $\epsilon_c$ of the
right-hand side, after that there is no solution for $\epsilon(f)$. Expanding
close to the maximum we obtain $f\simeq f_c+B(\epsilon-\epsilon_c)^2$, which
implies that the average rate of bond
failures increases very rapidly before fracture:
\begin{equation}
{d\epsilon \over df} \sim (f_c-f)^{-1/2}.
\label{eq:fbm_rho3}
\end{equation}
Contrary to the case of extreme type models, the strength of a fiber bundle
with ELS does not vanish in the large $N$ limit.
In particular, for any threshold distribution such
that $1-p(x)$ goes to zero faster than $1/x$ for $x\to \infty$, the
strength distribution is Gaussian with average $f_c=\epsilon_c(1-p(\epsilon_c))$ and
standard deviation $\sigma_f=\epsilon_c p(\epsilon_c)(1-p(\epsilon_c))/\sqrt{N}$
\cite{daniels45}.

A typical LLS model considers a one dimensional series of fibers
loaded in parallel with random breaking thresholds from a
distribution $p(x)$. When the load on a fiber exceeds the threshold
its load is redistributed to the neighboring intact fibers. Thus the
load on a fiber is given by $f_i= f(1+k/2)$, where $k$ is the number
of failed fibers that are nearest neighbors of the fiber $i$ and
$f=F/N$ is the external load \cite{harlow78}. Even for this
apparently
 simple one dimensional model a closed form solution is not
available, but several results are known from numerical simulations,
exact enumeration and approximate analytical methods
 \cite{harlow78,smith80,smith81,beyerlein96,phoenix97,phoenix92,curtin91,curtin93,zhou95,leath94,zhang96}
Contrary to the ELS model, LLS fiber bundles normally exhibit non-trivial
size effects as could be anticipated from general consideration of extreme
value statistics. In particular,  in the limit of large $N$, it has been shown that
the strength distribution follows the form
\begin{equation}
W(f)=1-[1-C(f)]^N \simeq 1-\exp(-N C(f)),
\label{eq:str_LLS}
\end{equation}
where $C(f)$ is a characteristic function, close to the Weibull
form, but difficult to determine exactly \cite{harlow78}. The existence
of a limit distribution has been recently proved under very generic
conditions for the disorder distribution in Ref.~\cite{mahesh04}, but
a numerical estimate of its form typically requires extremely large samples.
From Eq. \ref{eq:str_LLS} follows that the average bundle strength decreases
with $N$ as
\begin{equation}
f_c \sim 1/\log(N),
\end{equation}
so that an infinitely large bundle would have zero strength.

\section{Size effects from energy and geometry}
In the previous section, we discussed the statistical size effect,
which is caused by the randomness in the material strength. In this
section, we discuss two alternate ideas relevant, namely, the
energetic size effect and the geometric size effect on the failure
strength of quasi-brittle materials. The energetic size effect is
based on the Griffith's criterion of energy balance for stable crack
propagation. Intuitively, the energetic size effect arises due to
stress redistribution and the associated stored energy release as a
large fracture process zone (FPZ) develops ahead of the crack tips.
The size of the FPZ introduces an additional length scale into the
problem and its influence on the fracture strength is the energetic
size effect in quasi-brittle materials. This idea does not derive
from statistical fracture, but turns out to be quite useful in
understanding the results of the next section.

On the other hand, the geometric size effect is a consequence of
fractal geometry of crack surfaces. The morphology of crack surfaces
in disordered media can then be described by a self-affine
transformation, which implies a length-scale dependence of the
surface energy of a crack. This effect is supposed to introduce a
geometric size effect on the fracture strength.

\subsection{Energetic size effect}
Classical continuum theories of elasticity, plasticity and damage
mechanics do not possess a characteristic material length scale
$\ell$. Consequently, the nominal strength of geometrically similar
structures is independent structure size, i.e., there is no size
effect. Alternatively, the presence of a characteristic length scale
introduces a size effect on the nominal strength of the material.

Recall the basic idea of Griffith's criterion, that during
quasi-static crack propagation the available energy must be equal to
the energy required to create new crack surfaces. This is precisely
the expression derived in Eq.~(\ref{eq:griffith}) for brittle
materials under plane stress criterion. For quasi-brittle materials,
i.e., when the fracture process zone is comparable to or larger than
the structural size, Griffith's criterion can be generalized using
an effective crack size concept that accounts for the influence of
FPZ size on nominal strength.

Following the classical continuum fracture mechanics treatment, a generic expression for
the stress intensity factor around a crack of size $a$ may be expressed as
\begin{eqnarray}
K_I & = & \sigma \sqrt{\pi a} ~\kappa(\alpha) \label{eq:KI}
\end{eqnarray}
where $\kappa(\alpha)$ denotes a dimensionless function, $\alpha =
a/L$ denotes the relative crack size with $a = a_0 + \Delta a$ such
that $a_0$ is the initial crack size and $\Delta a$ is the
incremental crack size, and $L$ is a characteristic system size.
Equation~(\ref{eq:KI}) represents an extension of the classical
stress intensity factor around a central notch of size $2a$ in an
infinite panel ($K_I = \sigma \sqrt{\pi a}$ with $\kappa(\alpha) =
1$) to finite system sizes and various boundary conditions.
 The influence of characteristic system size $L$ on
stress intensity factor can be seen clearly by rewriting
Eq.~(\ref{eq:KI}) as
\begin{eqnarray}
K_I & = & \sigma \sqrt{L} ~\phi(\alpha) \label{eq:KI2}
\end{eqnarray}
where $\phi(\alpha) = \sqrt{\pi \alpha} \kappa(\alpha)$.

Using Irwin's relation, the generic expression for elastic energy release rate ${\cal G}$
can be written as
\begin{eqnarray}
{\cal G} & = & \frac{K_I^2}{E} = \frac{\sigma^2 L}{E} g(\alpha) \label{eq:G}
\end{eqnarray}
where $g(\alpha) = \phi^2(\alpha)$. For quasi-static crack growth, Griffith's criterion
states that
\begin{eqnarray}
{\cal G} & = & {\cal R} \label{eq:GR}
\end{eqnarray}
where ${\cal R}$ represents the crack resistance curve (R-curve: ${\cal R}(\Delta a)$), which
is expressed more generically as
\begin{eqnarray}
{\cal R} & = & R ~\psi(\alpha) \label{eq:R}.
\end{eqnarray}

As mentioned earlier, the fracture of quasi-brittle materials is
preceded by the development of fracture process zone, wherein the
material undergoes progressive damage due to microcracking and
interface failure. The presence of the  FPZ introduces a
characteristic length scale $\ell$ into the problem in addition to
the already existing length scales: $a_0$, $\Delta a$ and $L$. A
simple dimensional analysis dictates that the elastic energy release
rate and crack resistance curves be expressed in terms of
dimensionless ratios
\begin{eqnarray}
\eta & = & \frac{\Delta a}{\ell} ~~~~ \mbox{and} ~~~~ \vartheta = \frac{\ell}{L} \label{eq:eta}
\end{eqnarray}
in addition to the initial relative crack size $\alpha_0 = \frac{a_0}{L}$. Since
$\alpha = \alpha_0 + \eta \vartheta$, the energy release rate and the crack resistance
can be expressed as
\begin{eqnarray}
{\cal G} & = & \frac{\sigma^2 L}{E} g(\alpha_0, \eta, \vartheta) \\
{\cal R} & = & R \psi(\alpha_0, \eta, \vartheta)´.
\end{eqnarray}

For fracture to occur at the peak load, we require that
\begin{eqnarray}
\frac{\partial {\cal G}}{\partial \Delta a} & = & \frac{\partial
{\cal R}}{\partial \Delta a}. \label{eq:peak}
\end{eqnarray}
Using the relations $G = R$ and $\eta = \frac{\Delta a}{L}$,
Eq.~(\ref{eq:peak}) may alternatively be expressed as
\begin{eqnarray}
\frac{\partial \phi(\alpha_0, \eta, \vartheta)}{\partial \eta} & = &
\frac{\partial \psi(\alpha_0, \eta, \vartheta)}{\partial \eta}.
\label{eq:peak1}
\end{eqnarray}
From Eq.~(\ref{eq:peak1}), one can derive an implicit expression for
the critical crack as
\begin{eqnarray}
\eta & = & \eta_c(\alpha_0, \vartheta). \label{eq:etac}
\end{eqnarray}
Substituting Eq.~(\ref{eq:etac}) into Griffith's fracture criterion
${\cal G} = {\cal R}$ and simplifying the result, we get
\begin{eqnarray}
\sigma_c & = & \frac{\sqrt{E G}}{\sqrt{L h(\alpha_0, \vartheta)}} \label{eq:sigma}
\end{eqnarray}
where
\begin{eqnarray}
h(\alpha_0, \vartheta) & = & \frac{g(\alpha_0, \eta_c(\alpha_0,
\vartheta), \vartheta)}{\psi(\alpha_0, \eta_c(\alpha_0, \vartheta),
\vartheta)}. \label{eq:h}
\end{eqnarray}

For very large system sizes $L$, $\vartheta$ approaches zero
asymptotically, i.e., $\vartheta \rightarrow 0$. Hence, $h(\alpha_0,
\vartheta)$ can be approximated as
\begin{eqnarray}
h(\alpha_0, \vartheta) & \approx & h_0(\alpha_0) + h_1 \vartheta \label{eq:hexpand}
\end{eqnarray}
where
\begin{eqnarray}
h_0 & \equiv & h(\alpha_0,0) \\
h_1 & \equiv & \frac{\partial h(\alpha_0,\vartheta)}{\partial
\vartheta}|_{\vartheta = 0}.
\end{eqnarray}
Substituting Eq.~(\ref{eq:hexpand}) into Eq.~(\ref{eq:sigma}), we
have
\begin{eqnarray}
\sigma_c & = & \frac{\sqrt{E G}}{\sqrt{L h_0 + \ell h_1}}.
\label{eq:sigmac}
\end{eqnarray}

For quasi-brittle materials with an initial notch and constant R-curve, we have
$h_0 = g(\alpha_0) \sim \alpha_0 = a_0/L$. Denoting $\xi = \ell h_1$, Eq. \ref{eq:sigmac}
can be specialized to the initially notched samples as
\begin{eqnarray}
\sigma_c & = & \frac{K_c}{\sqrt{a_0 + \xi}} \label{eq:sigmac1}.
\end{eqnarray}
Note that this argument states that $\xi$ depends on the partial
derivative of the ratio of $\cal G$ and $\cal R$ on $\vartheta$, so
it is related on the relative scales of  the two rates.

\subsection{Geometric size effect}
In this section, we discuss the implication of roughness of crack
surfaces on the linear elastic fracture mechanics and therewith on
the scaling of strength. For many years now, it is hoped that there
exists a simple relation between material toughness and the
self-affine exponent of rough cracks, although such a hope appears
too optimistic now (see Ref. \cite{bazant05} for controversies
related to this topic). An exception is the indirect fact that
explaining crack roughness via the theory of depinning of elastic
manifolds implies that the critical stress intensity factor can
be expressed via a few relevant parameters such as crack front
elasticity and the strength of the disorder, while the precise 
value depends on the geometry (planar cracks, notches or cracks in two or three
dimensions...) \cite{ramanathan97b,charles04,bonamy06,bonamy08}.
There is, however, no size effect except for the one that arises from
the finite crack length and the resulting correction to the critical
stress intensity factor, in analogy with finite size corrections
observed in critical phenomena \cite{charles04,alava06}. It is
important to note that the roughness and strength can be directly
coupled, since e.g. the crack growth resistance and the roughness exponent
(fractal dimension) follow from the same process. In what follows,
the empirical observations of strength are explained a posteriori
using as input a given roughness exponent for the crack.

In particular, roughness of crack surfaces influences both (i) the
stress concentration around the crack
\cite{mosolov93,balankin97,yavari02} and (ii) the energy required to
create crack opening through its dependence on the actual crack area
\cite{borodich97}. Experiments on several materials under different
loading conditions suggest that the crack surfaces exhibit
self-affine scaling, which implies that if the in plane length
scales of a fracture surface are scaled by a factor $\lambda$ then
the out of plane length scales (height) of the fracture surface
scales by $\lambda^\zeta$, where $\zeta$ is the roughness exponent.
This also implies that the actual (curvilinear) length of the crack
is a non-trivial function of length-scale. Consequently, an
infinitesimal crack opening will now cost {\em more surface energy}
than in the usual case, and thus the Griffith's criterion is
modified in a way to account for the increased energy requirement to
propagate cracks, and possibly to account for the modified stress
fields and hence the elastic energy due to a rough crack.

Following Refs. \cite{mosolov93,balankin97,yavari02}, the asymptotic near-tip
stress fields around a rough crack may be expressed as
\begin{eqnarray}
\sigma_{ij} & = & K_I r^{-\beta} f(\theta), \label{eq:roughsigma}
\end{eqnarray}
where $r$ is the radial distance from the crack tip, $f(\theta)$
denotes the functional dependence of stress concentration around
crack tip, and $\beta = 1 - 1/2\zeta$ for $0.5 \le \zeta \le 1$ and
$\beta = 0$ otherwise. Equation (\ref{eq:roughsigma}) implies that
the stress concentration around a rough crack can be expressed as
\cite{yavari02}
\begin{eqnarray}
K_I & \sim & \sigma a^{\beta} \chi(\beta) \label{eq:KIrough}
\end{eqnarray}
where $\chi(\beta)$ is a dimensionless function. The influence of
characteristic system size $L$ on stress intensity factor is
incorporated into Eq. (\ref{eq:KIrough}) as
\begin{eqnarray}
K_I & = & \sigma L^\beta ~\phi(\alpha) \label{eq:KI2rough} \\
G & = & \frac{\sigma^2 L^{2\beta}}{E} g(\alpha_0, \eta, \vartheta).
\label{eq:Grough}
\end{eqnarray}

From the crack resistance point of view, it is expected that the roughness of a
self-affine crack leads to R-curve behavior. Specifically, for a self-affine crack,
the crack surface energy scales as
\begin{eqnarray}
{\cal E}_{surf}  & = & 2 \gamma a^{1/\zeta} \label{eq:Esurf}
\end{eqnarray}
where $\gamma$ denotes the specific energy per
unit of fractal measure and $1/\zeta$ denotes the fractal dimension of the crack.
Hence, the crack resistance is given by
\begin{eqnarray}
R  & = & 2 \gamma a^{1/\zeta - 1} \label{eq:Rfractal}
\end{eqnarray}

We also note that a size effect law based on anomalous scaling of
crack surfaces has been proposed in Refs. \cite{morel00,morel02}. In
the case of anomalous scaling of crack surfaces (see
Eq.~(\ref{bouchsca}) below), the extra scale also enters the scaling
argument, further changing the amount of surface energy needed for
infinitesimal crack advancement. This implies a non-trivial
``R-curve'', or crack resistance, since the energy consumed via
forming new crack surface depends not only on the full set of
exponents but again on the distance propagated. An expression can be
written down using solely geometric arguments and the exponents,
such as

\begin{eqnarray*}
        \Delta h(\ell,x)\simeq A
        \left\{ \begin{array}{ll}
        \ell^{{\zeta}_{loc}} \: \xi (x)^{\zeta -{\zeta}_{loc}}
        & \mbox{if \quad $\ell \ll \xi (x)$} \\
        \xi (x)^{\zeta} & \mbox{if \quad $\ell \gg \xi (x)$}
\label{bouchsca}
          \end{array}
\right.
\end{eqnarray*}
where $\Delta h(\ell,x)$ denotes the height fluctuations of fracture
surfaces estimated over a window of size $\ell$ along the x-axis and
at a distance $x$ from the initial notch. Also, $\xi(x)=Bx^{1/z}$ is
a crossover length along the x-axis below which the fracture surface
is self-affine with local roughness exponent $\zeta_{loc}$. This
crossover length  depends on the distance $x$ to the initial notch.
For empirical reasons, one needs to note that there should be also a
lower cut-off above which the self-affinity can be observed
\cite{bouchaud94}, and that the argumentation should change if the
crack geometry is very branched \cite{bouchaud93}.

There are two different regimes: for $\ell \gg \xi (x)$, $\Delta
h(l,x) \sim x^{\zeta/z}$, and for $\ell \ll \xi (x)$, it is
characterized by the exponent $(\zeta-{\zeta}_{loc})/z$, where $z$
is the dynamic exponent. Since the effective area is (from which
surface energy follows)
\begin{equation}
    G \ \delta A_p = 2\gamma\ \delta A_r \label{eq:Gano}
\label{Crit}
\end{equation}
where $\delta A_r$ is the real area increment and $\delta A_p$ its
projection on the fracture mean plane. Morel et al. make now the
simplest first order approximation, that Eq.~(\ref{eq:Gano}) can be
approximated as $G = 2 \gamma s(x)/L$, where $s(x)$ is the length of
the crack profile that can be estimated by covering the profile path
with segments of length $\delta$ whose horizontal projection on
x-axis is $\ell_0$. The elastic energy released is not taken to be
affected by the self-affinity of the crack. Consequently, we have
$s(x) = (L/\ell_0) \delta$, which can be approximated as
\begin{eqnarray*}
        s(x) \simeq L
        \left\{ \begin{array}{ll}
        \left[1 + \left(\frac{A (B x^{1/z})^{\zeta-\zeta_{loc}}}{\ell_0^{1-\zeta_{loc}}}\right)^2\right]^{1/2}
        & \mbox{if \quad $x \ll x_{sat}$} \\
        \left[1 + \left(\frac{A L^{\zeta-\zeta_{loc}}}{\ell_0^{1-\zeta_{loc}}}\right)^2\right]^{1/2} & \mbox{if \quad $x \gg x_{sat}$}
\label{ssca}
          \end{array}
\right.
\end{eqnarray*}

Substituting Eq. \ref{ssca} into $G = 2 \gamma s(x)/L$, we get an expression for
crack resistance as
\begin{eqnarray*}
        G_R(\Delta a \ll \Delta a_{sat}) & \simeq &
2 \gamma
        \sqrt{1+\left( \frac{A B^{\zeta-{\zeta}_{loc}}}
        {{l_o}^{1-{\zeta}_{loc}}} \right)^2
        {\Delta a}^{2(\zeta-{\zeta}_{loc})/z}}
\label{Gr}
\end{eqnarray*}
within the region of crack growth ($\Delta a \ll \Delta a_{sat} = x_{sat}$), and
when $\Delta a \gg \Delta a_{sat}$, the crack growth resistance
becomes
\begin{eqnarray*}
        G_{R}(\Delta a \gg {\Delta a}_{sat}) &\simeq &
2 \gamma
        \sqrt{1+\left( \frac{A}
        {{l_o}^{1-{\zeta}_{loc}}} \right)^2
        L^{2(\zeta-{\zeta}_{loc})}}.
\end{eqnarray*}

\section{Size effects from disordered fracture models}

Simulations of statistical fracture provide a tool to analyze the
validity of various scenarios for the size-dependence of strength.
The major tool for this  has been the Random Fuse Model
\cite{deArcangelis85,alava06}, where one approximates continuum
Linear Elastic Fracture Mechanics (LEFM) with a spatial
discretization (lattice) and a scalar analogy (voltages) for the
displacements. The material response is formulated in terms of "fuse
elements" which have a conductivity $G_0$ and a failure threshold
$i_c$. Usually the "elasticity" or the fuse conductances is kept
constant, while various threshold distributions $p(i_c)$
are employed. Until now almost all the effort has been
spent on two-dimensional systems for the sake of numerical
convenience. The discussion of established results can be divided
into two topics, studies of the weakest-link effect and analyses of
the size effect for notched specimens.

\subsection{Statistical size effect}
The major results on the strength of the RFM were obtained in the
late 80's by Duxbury and co-workers \cite{duxbury87,duxbury88}. The
starting point is to consider {\em weak disorder} employing a
failure threshold distribution composed by two delta-functions at
$i_c=0$ and $i_c=1$ which naturally translates into "dilution
disorder" or the removal of a fraction $1-p$ of the fuses at the
beginning. In the diluted limit ($p \simeq 1$), it is possible to
formulate a non-rigorous theory of the size-effect. First, the flaw
size statistics indicates an exponentially decaying length
distribution for the microcracks that are induced into the samples
by the dilution disorder. Second, in practice the critical current
$I_c$ essentially corresponds to the current $I_1$ st which the
weakest fuse fails. This means that the case under study is very
brittle, and the crack growth resistance is small. Given these two
assumptions, Duxbury {\em et al.} write down a modified extremal
statistics theory of the type of Eq. (\ref{eq:gumbel}), and compare
it to RFM simulations in two dimensions with small dilution.
There are straightforward generalizations of the weak-disorder
dilution theory towards $p\sim p_c$, the bond percolation of the RFM
geometry (for square lattices, $p_c=1/2$), since at the percolation
point the outcome can be simply understood via the "red bond" or
singly-connected bond arguments of percolation theory
\cite{duxbury88}. This limit is not relevant for most materials, but
in general the defect size statistics can not be assumed
exponential, and often a noticeable damage accumulation takes place
before failure.

\begin{figure}[t]
\begin{center}
\includegraphics[width=10cm]{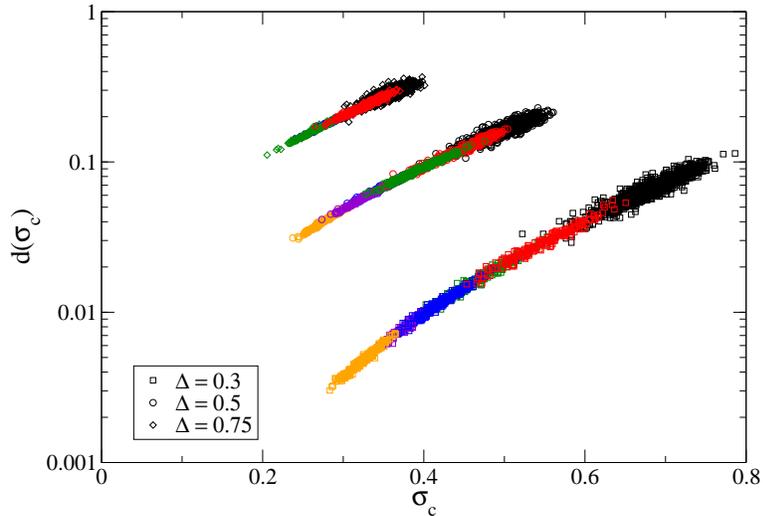}
\end{center}
\caption{The accumulated damage $d$ as a function of the failure stress $\sigma_c$ on a sample-to-sample
 basis for three different disorder strengths $\Delta$ and several lattice sizes $L$.}
\label{figDamStre}
\end{figure}

To this end, we consider a variant of the RFM where $p(i_c)$ has a
finite support which  extends down to zero. In particular, we
consider disorder with a cumulative distribution $P(i_c) =
i_c^\Delta$, $i_c \in [0,1]$. The case $\Delta=1$ corresponds to the
uniform distribution, while different values of $\Delta$ allow to
tune the strength of disorder. Since a fuse can fail even at low
currents, with these kind of distributions there is a considerable
damage accumulation before failure. We illustrate this in Fig.
\ref{figDamStre} which shows the relation between the accumulated
damage $d$ and the failure stress $\sigma_c\equiv I_c/L$ The figure
shows the results of two dimensional simulations for three different
disorder strengths $\Delta$ and for several system sizes from $L=64$
to $L=320$ or even $512$ (for the smallest $\Delta$ or weakest
disorder). It is interesting to note that the $d(\sigma_c)$ relation
follows roughly a power-law. While this general behavior could be
anticipated on the basis of a simple mean-field relation
$d=\int_0^{\sigma_c} di_c p(i_c)= \sigma_c^\Delta$, the measured
curves do not follow this prediction. As we discuss below in more
detail, the damage is a sum of a statistically homogeneous
background plus the FPZ contribution so the failure of the
mean-field is not a surprise.

\begin{figure}[t]
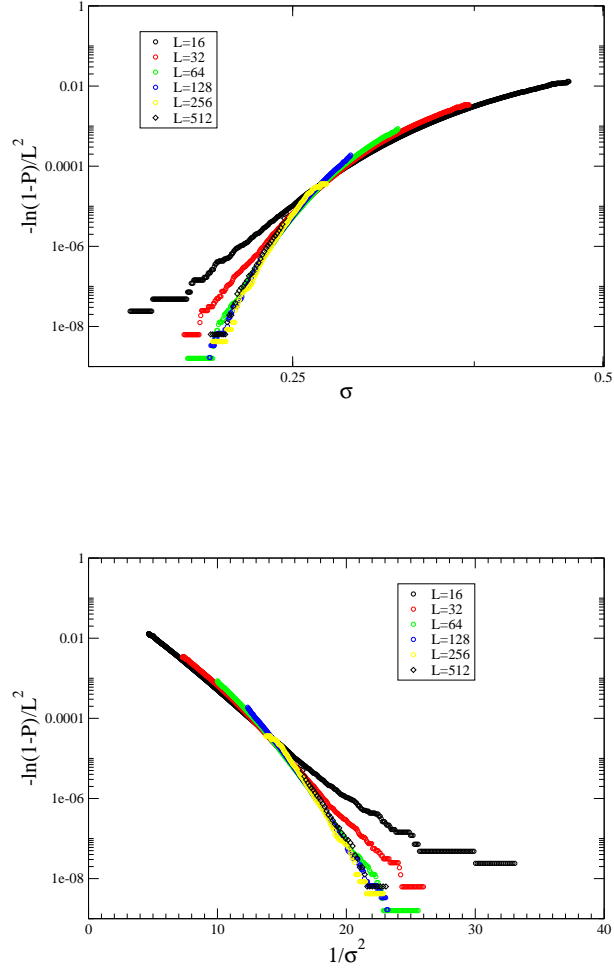

\begin{center}
\includegraphics[width=8cm]{./weibull.eps}

\vspace{2cm}

\includegraphics[width=8cm]{./duxbury1.eps}

\vspace{1cm}
\end{center}
\caption{Cumulative strength distributions for the RFM (with $\Delta=0.6$) plotted for several $L$.
 Upper panel: the data shown using a ``Weibull paper''. If the data would obey the Weibull
distribution, they should all collapse into a single a straight
line. Lower panel: a  generalized ``Gumbel'' plot. Again a collapse
to a straight line would be required if the data were described by a
generalized Gumbel distribution.} \label{figGumbelOwn}
\end{figure}

Damage accumulation is responsible for a fundamental complication in
the evaluation of size effects which should be relevant for real
materials. To relate this to the discussion in Section 2, the
cumulative $P(a)$ in Eq.~(\ref{eq:pna}) becomes dependent on
$\sigma$. Fig. \ref{figGumbelOwn} shows numerical results for the
cumulative strength distribution $P(\sigma_c,L)$ for the RFM. As
demonstrated in the Figure, the data are not described by a Weibull
distribution, but not even the modified Gumbel distribution
mentioned above captures the observed behavior. It is interesting
to note nevertheless, that for larger $L$ the double exponential
plot seems to reveal the presence of an asymptotic scaling form.
More importantly, the behavior here shows that the results are not
easily interpreted in terms of extreme statistics in its simples
forms. It would be desirable to formulate a theory capturing the
evolution of $P(i_c,L)$ as $L$ increases. Studies of idealized
chains-of-fiber-bundle models have indicated the presence of
cross-over effects depending on system size and disorder and the
failure of standard extremal statistics scalings
\cite{mahesh99,duxbury94,mahesh02}. Such chains consist of simple
one- or two-dimensional models in series. Thus their behavior is
dictated when $V$ changes, in analogy to Sec.~\ref{sec:weak},
by the chain length $N$ and the distribution $P(\sigma)$. Important
physics originates at the low-strength tails of the element (fiber,
fuse, bond, element) strength distribution and from the mesoscopic
damage dynamics. Other important effects are crack arrest by strong
elements, and for weaker disorder the concept of a critical defect
size, reaching which anywhere in the sample leads to catastrophic
failure.

\subsection{Strength of materials with flaws}

Another fundamental question is whether Eq.~(\ref{eq:griffith}) from
the Griffiths' argument remains valid for realistic materials. A
suitable generalization as discussed in the previous section is $
\sigma_c = K_c/\sqrt{\xi + a_0}$, where we have added a scale $\xi$
and written  $K_c \sim \sqrt{E G_c}$ for the fracture toughness
(numerical factors relating eg. plane stress or plane strain loading
scenarios have been omitted). $\xi$ plays here the role of
incorporating a disorder effect: that cracks are masked when they
become very small by fluctuations, and thus the strength saturates
to $K_c/\sqrt{\xi}$. As noted by Bazant (see e.g. \cite{bazant04b})
a natural reason for the scale-length is the Fracture Process Zone
which is non-negligible in quasi-brittle media.

\begin{figure}[ht]
\begin{center}
\vspace{1cm}
\includegraphics[width=8cm]{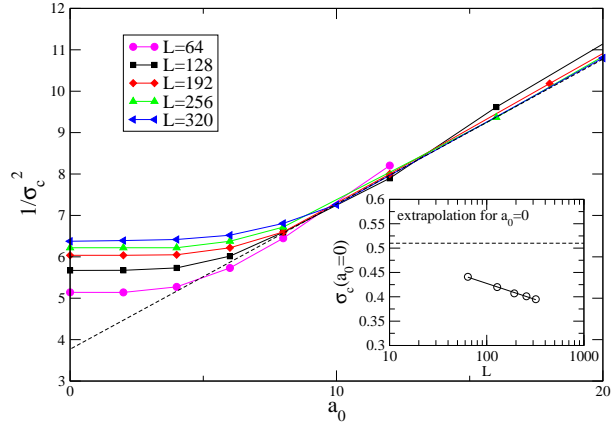}\end{center}
\caption{RFM strength data for intermediate disorder and varying
flaw sizes $a_0$ and sample linear sizes $L$. The inset compares
fracture strength in unnotched samples with that predicted by
Eq.~(\ref{eq:sigmac1}) for $a_0=0$.} \label{figBazant}
\end{figure}

Extensive numerical simulations of the RFM and also other models -
Random Spring Model, Random Beam Model - have allowed to understand
the physics of size effect here and to elaborate on a scaling
theory \cite{alava08a,alava08b}. Figure \ref{figBazant} shows an example of 2d RFM data
plotted using the inverse square of Eq.~(\ref{eq:sigmac1}). For
small flaws one sees a cross-over away from the behavior predicted
by the above scaling, and the inset illustrates how this depends on
the specimen size, which is otherwise absent in principle from
Eq.~(\ref{eq:sigmac1}).

The dependence of the size-effect on $L$ and the disorder strength
$\Delta$ (in full generality, we take $\Delta=0$ to correspond to pure systems
with no disorder) can be summarized with the scaling
\begin{equation}
\frac{K_c^2}{\sigma_c^2} = \xi+a_0 f(a_c/a_0) \label{fundamental}
\end{equation}
where the scaling function $f(y)$ follows
\begin{eqnarray}
        f(y)\simeq
        \left\{ \begin{array}{ll}
        1  & \mbox{if \quad $y \ll 1$}  \\
        y & \mbox{if \quad $y \gg 1$}
\label{scafunc}
                \end{array}
\right. \label{scafu}
\end{eqnarray}
The important point here is that Eq.~(\ref{scafu}) takes into
account of the weak-link effects discussed above. The length-scale
$a_c \simeq (K_c(\Delta) /\sigma(L,\Delta))^2-\xi(\Delta)$ indicates the cross-over
between that regime and the LEFM one where Eq.~(\ref{eq:sigmac1}) is
valid.

\begin{figure}[t]
\begin{center}
\includegraphics[width=8cm]{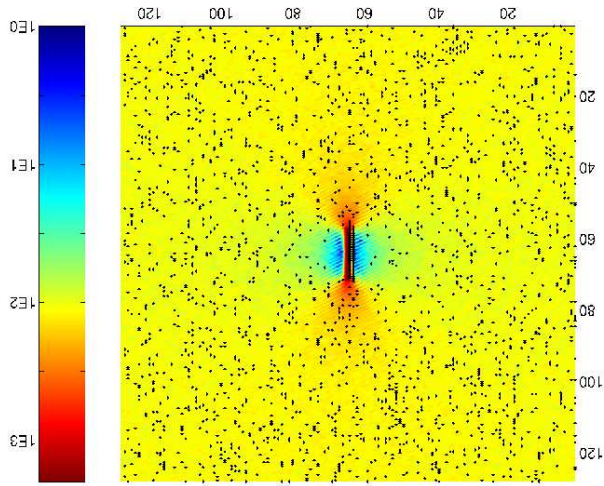}
\end{center}
\caption{The damage at maximum stress $\sigma_c$ for a RFM with
linear size $L=128$ and intermediate disorder $\Delta=0.6$. The black
markers illustrate the broken fuses in a single sample, and the
color code depicts the average damage over $N=2000$ samples. One can
see screening effects next to the notch boundaries (weak damage),
and then two clouds of damage next to the notch tips.} \label{xifig}
\end{figure}

The size-effect can thus be summarized with the aid of two
parameters, $\xi$ and $K_c = \sqrt{GE}$ where $E$ is the elastic
modulus, and $G$ the fracture energy. It has to be emphasized that
all the three quantities of $\xi$, $E$, and $G$ depend on the
presence of disorder. Figure \ref{xifig} illustrates the FPZ damage
profile, which underlies $\xi$. It measures the exponential decay of
the damage from the crack tip, in a situation where the LEFM stress
profile (and its correspondence in the RFM) is not seen in the
damage due to screening. The stronger is the disorder, the larger is
$\xi$ as one would expect, indicating also a strength reduction. It
would be very interesting to develop an analytical model of this
screening effect. The figure \ref{xifig} demonstrates however that
there are strong sample-to-sample fluctuations: the exponential
damage profile and its role in the average strength behavior or
size-effect is to be understood as a statistical average.

It is interesting to note that these statistical models exhibit a
non-trivial R-curve if one computes the crack extension using the
changing scale of the damage cloud, $\xi(\sigma)$. The strength
prediction of Eq.~(\ref{eq:sigmac1}) is then ``self-consistent''.
For those $a_0$ for which it is valid, $\xi$ is a constant to which
value $\xi(\sigma)$ grows as damage accumulates from zero. Likewise,
$K_c$ is to be measured at $\sigma_c$ as well. There are no clear
signs of any kind of geometrical size effects, since though the
cracks formed are self-affine, the growth takes place after the
stress maximum \cite{alava08b}. However, the damage cloud is truly
two-dimensional and we thus note that the results differ from those
obtained for uniaxial fiber composites
\cite{beyerlein99a,beyerlein99b} which exhibit a scaling similar to
Eq.~(\ref{eq:sigmac1}) but with a logarithmic correction and which
reproduce a R-curve in a similar fashion as the statistical models.

\begin{figure}[t]
\begin{center}
\includegraphics[width=8cm]{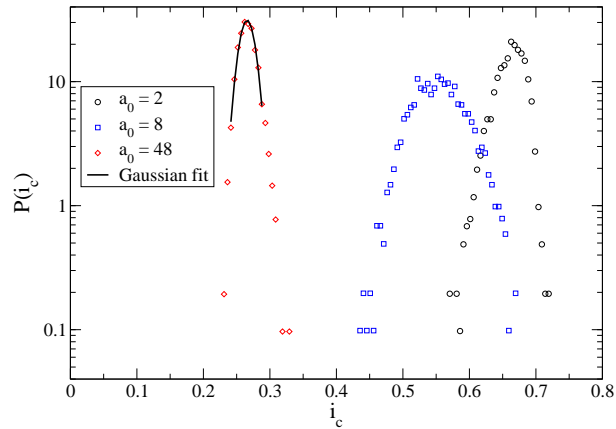}
\end{center}
\caption{The distribution $p(\sigma_c,a_0)$ for a RFM with linear
size $L=192$ and intermediate disorder $\Delta=0.6$. Three different
$a_0$ are used: $L=4, 16, 48$. The last dataset is compared to a
Gaussian fit in the central part around $\sigma_c -   \langle
\sigma_c \rangle$. The numerical distribution are obtained from 1000
to 4000 samples depending on $\Delta$ and $a_0$. } \label{notchdist}
\end{figure}

This brings us to the important issue of strength fluctuations in
the presence of a dominating notch. In the previous sections we have
discussed the important role of extremal statistics in this. Now,
the size-effect, of the whole distribution. is not expressed as a function of $L$ but of $a_0$.
Figure \ref{notchdist} illustrates the behavior of
$P(\sigma_c,a_0)$. As $a_0$ grows the average strength goes down,
and the width of the distribution reaches a maximum at an intermediate value
of $a_0$. For small notches, $a_0 < a_c$ the distribution naturally
is close to the finite-size form for $p(\sigma_c,a_0 = 0,L)$. For large
values of $a_0$ we see a narrowing of the distribution (in fact,
$\Delta \sigma_c / \sigma_c$ goes to zero with $a_0$ increasing) and
the distribution is well approximated in the central part by a Gaussian
shape. This is interesting, since the Gaussian distribution is also
obtained in global load sharing fiber bundles
\cite{daniels45} and in chains-of-fiber-bundle
models with local load sharing and heterogeneous fiber strengths
\cite{mahesh02}. However, even for the largest $a_0 = 48$ the
tails are broader than in the Gaussian case. It would perhaps
be expected that for the weak strength limit the fuse threshold
distribution would be relevant \cite{mahesh99}, and for this reason
it would be interesting to study these models also with e.g. a Weibull $P(i_c)$.

\section{Conclusions}
In this paper we have reviewed various statistical approaches to
fracture size effect. We first discussed the Weibull theory and
similar approaches based on the weakest link concept. According to
these extreme value statistics arguments, larger samples should be
more likely to possess weak regions and are thus bound to fail at a
smaller load. While this general statistical argument is generally
believed to be correct, the presence of the stress enhancement
around a crack and damage accumulation prior to failure make a
quantitative theory for size effects in materials still elusive. In
cases when a dominating crack, or a notch, is present one can rely
on energetic arguments that link the crack geometry to the failure
stress. Further complications arise when the dominating crack is not
straight but has a self-affine geometry, as observed in experiments.

These general statistical approaches have been tested using
statistical lattice models for fracture, where a set of discrete
elements with random failure thresholds are subjected to an
increasing load. These models allow to study the crossover between
notch dominated size-effects and statistically induced ones. Despite
this, the role of damage accumulation for fracture size effects in
unnotched samples still remains unclear. On the experimental side,
large statistical sampling would be needed to reach firm conclusions
about the asymptotic behavior of materials strength. The above
observations apply for quasi-brittle fracture first and foremost,
and if viscoelastic or viscoplastic deformation is important our
understanding is much less developed. These important questions
remain open for future investigation, so that statistical fracture
promise to pose interesting challenges for the years to come.

{\bf Acknowledgments -} MJA would like to acknowledge the support of
the Center of Excellence -program of the Academy of Finland.
MJA and SZ gratefully thank the financial support
of the European Commissions NEST Pathfinder programme TRIGS under
contract NEST-2005-PATH-COM-043386. PKKVN acknowledges support from
Mathematical, Information and Computational Sciences Division,
Office of Advanced Scientific Computing Research, U.S. Department of
Energy under contract number DE-AC05-00OR22725 with UT-Battelle,
LLC.

\providecommand{\newblock}{}


\end{document}